\documentclass[a4paper,11pt]{article}
\pdfoutput=1 % if your are submitting a pdflatex (i.e. if you have
\usepackage{xspace}
\usepackage{subcaption}
\usepackage{mathrsfs} 
\usepackage{xcolor}
\usepackage{dsfont}
\usepackage{xspace}
\usepackage{tabularx}

\usepackage{comment}

\usepackage{jheppub} % for details on the use of the package, please
\newcommand{\df}{\mathrm{d}}
\newcommand{\img}{\mathrm{i}}

\newcommand{\deco}{\texttt{DECO}\xspace}

\preprint{Nikhef 2022-018}

\title{\boldmath Discrete symmetries and Efficient Counting of Operators}

\author[a,b]{Simon Calò,}
\author[a,b]{Coenraad Marinissen,}
\author[a,b,c]{Rudi Rahn}

\affiliation[a]{Nikhef, Theory Group,
	Science Park 105, 1098 XG, Amsterdam, The Netherlands}
\affiliation[b]{Institute for Theoretical Physics Amsterdam and Delta Institute for Theoretical Physics, University of Amsterdam, Science Park 904, 1098 XH Amsterdam, The Netherlands}
\affiliation[c]{Department of Physics and Astronomy, University of Manchester, Manchester, M13 9PL,United
Kingdom}

\emailAdd{simon.calo@student.uva.nl}
\emailAdd{c.b.marinissen@uva.nl}
\emailAdd{rudi.rahn@manchester.ac.uk}

\abstract{We present \deco (``Discrete and Efficient Counting of Operators''), an implementation of the Hilbert Series to enumerate subleading operator bases for SMEFT-like EFTs with symmetry groups as typically found in flavour and BSM physics. \deco can accommodate EFTs with arbitrary numbers and combinations of the SM gauge groups, as well as the discrete groups $S_4$, $A_4$, and $\mathds{Z}_n$, and U(1) groups with residual global charge (and these groups' most important representations). The program is highly modular and can easily be extended to additional groups and/or representations. We demonstrate the design cases for \deco by using it to cross-check subleading operator bases of EFTs in the literature, which allows us to identify a missing operator in a widely used model for the neutrino masses and discuss said operator's impact. }

\begin{document} 
\maketitle
\flushbottom

\section{Introduction}
\label{sec:intro}
Effective Field Theories (EFTs)~\cite{Manohar:2018aog,Cohen:2019wxr} offer a model-independent way of including the effects of unknown high-energy physics on the dynamics of low-energy fields, are an invaluable tool in the absence of confirmed discoveries of physics beyond the Standard Model (BSM), and offer one of the most flexible approaches to model any New Physics. In a hybrid regime EFTs can be used to study extensions of the Standard Model (SM) in which the SM fields and charges are supplemented with additional fields or symmetries, to study the effects of fields just out of reach of direct discovery.

The latter is an approach employed for example in flavour physics, in the hope that the imprint of some ordering principle behind the generational pattern of masses and couplings can be discerned: The SM is extended with some new field or some new symmetry in the hope to reduce the typically large number of subleading operators by promoting apparent patterns to features of the model (see e.g.~\cite{Bordone:2019uzc,Feruglio:2019ybq,Faroughy:2020ina,Greljo:2022cah}).

A certain temptation then presents itself: We wish to devote our time \emph{studying} subleading operator bases, but not to construct and especially \emph{verify} them, leading to the risk of working with over- or undercomplete bases. A helpful tool in this endeavour is the Hilbert series~\cite{Lehman:2015via,Lehman:2015coa,Henning:2015daa}, which can be used to derive the number and field content (though not the precise index structure) of effective operators. Remarkable progress has been made in recent years to apply the Hilbert Series to both weakly and strongly coupled EFTs~\cite{Henning:2015alf,Graf:2020yxt}, including complications from charge conjugation and CP transformations~\cite{Henning:2017fpj,Graf:2020yxt,Kondo:2022wcw,Sun:2022aag}, and to derive insight into the structure of EFTs at high~\cite{Melia:2020pzd}, as well as all~\cite{Helset:2020yio,Talbert:2022unj} effective orders. Recently, the application of the Hilbert Series to EFTs is also beginning to intersect with some earlier applications, like the study of flavour invariants in EFTs~\cite{Yu:2022nxj,Yu:2022ttm}. A number of computational tools has been released to assist with various such tasks~\cite{Macaulay,Henning:2015alf,Gripaios:2018zrz,Criado:2019ugp,Marinissen:2020jmb,Banerjee:2020bym,Kondo:2022wcw}. However, the set of symmetry groups considered in the Hilbert series context --- especially when it comes to computational tools --- is still mostly limited to the continuous groups encountered as gauge redundancies and accidental symmetries of the SM, and in particular does not extend to the large set of discrete groups employed in e.g. flavour physics. Flavour symmetric models constitute a large playground for Hilbert Series techniques, simply because there are a lot of possibilities:
As there is a large number of discrete groups with appropriate representations (see e.g.~\cite{Ishimori:2010au} for appearing mathematical groups), and the groups can be freely combined, the number of potential symmetries to be present is vast (see~\cite{Parattu:2010cy} for an impression of the sheer number of potential groups). A large body of literature exists applying these ideas to both lepton and quark sectors already (see e.g.~\cite{Hernandez:2012ra,Lam:2012ga,Holthausen:2012wt,King:2013vna,Holthausen:2013vba,Joshipura:2014pqa,Talbert:2014bda,Yao:2015dwa,King:2016pgv,deMedeirosVarzielas:2016fqq,Yao:2016zev,Bernigaud:2019bfy}), and research involving discrete symmetry groups is still actively ongoing (see~\cite{Bernigaud:2022sgk,Garg:2022nor,Baur:2022hma,Vien:2022sxh,Vien:2022pwf,Kobayashi:2022jvy,Gautam:2022jrg,Puyam:2022mej,KrishnanR:2022rtm,Singh:2022tge,deAnda:2021jzc,Nomura:2021ewm,Bonilla:2021ize,Hagedorn:2021ldq,Srivastava:2021pyz,Hernandez:2021mxo,Verma:2021koo,Nguyen:2020ehj,deMedeirosVarzielas:2021pug,Nomura:2022boj,Ghosh:2021rmn} for examples from 2022 alone). We thus aim to lay the groundwork here to chart this so far unexplored aspect of EFT applications of the Hilbert series.

In this paper we present a modular approach to the Hilbert series for weakly coupled EFTs via a \texttt{FORM}~\cite{Vermaseren:2000nd,Ruijl:2017dtg} program called \deco (``Discrete and Efficient Counting of Operators'', published at \url{https://www.github.com/cbmarini/deco}), allowing a user to swiftly derive the number of subleading operators at some EFT order (given by the operators' mass dimension) for EFTs with freely chosen field content and charge pattern. ``Freely chosen'' here refers to two levels of modularity: First, we allow for arbitrary numbers and types of fields with ``standard'' Lorentz spin (i.e. scalars, chiral fermions, field strength tensors, Weyl tensors), and arbitrary number and type of symmetry charges under a set of internal symmetry groups which includes the usual SM-like symmetries. And second, we keep our code flexible enough to easily allow us to add additional representations and symmetry groups. In other words and to exemplify: we can accommodate support for a hypothetical model with 5 independent lefthanded fermions charged under various representations of 12 different SU(3) symmetries already, and can easily add support for e.g. a $\Delta(75)$ flavour symmetry upon request.
In particular \deco currently offers support --- beyond the typical Standard Model Lie groups SU(3), SU(2), and U(1) --- for the discrete groups $A_4$ and $S_4$, as often encountered in flavour physics (see e.g. the more than 50 models listed in~\cite{Altarelli:2010gt}, and more recent reviews~\cite{Petcov:2017ggy,Petcov:2018snn}), as well as $\mathds{Z}_n$ groups of arbitrary cycle length $n$ and U(1) groups with residual charge (i.e. U(1) groups under which all effective operators carry a defined non-zero charge), as encountered e.g. in the low-energy implementation of supersymmetric models (e.g.~\cite{Altarelli:2005yx,Altarelli:2008bg}).

The paper is structured as follows: In section 2 we recap the mathematical background of the Hilbert series to the extent necessary to understand how discrete symmetry groups fit into the approach. In section 3 we comment on the HS for discrete groups, and derive the knowledge and computational input required for the implementation of the groups $A_4$ and $S_4$. In section 4 we elaborate on the technical aspects of our implementation and the program \deco. In section 5 we present application cases, including one in which we use \deco to track down an operator overlooked in a template model for flavour symmetry, and comment on the implications of this missing operator. Finally, we conclude in section 6.

\section{Counting Effective Operators}
\label{sec:hilbert}

A crucial aspect of any study of higher order effects in an EFT arises from the
requirement that the set of subleading operators has to be minimal, to avoid
double-counting the effects of the physics that is integrated out. Traditionally
this is achieved by brute force, i.e. by exploiting symmetry relations and
matrix identities to show that operators are related, and subsequently removing
elements of the set of operators. This is obviously error-prone and tedious. A
useful tool to assist in this endeavour presents itself in the form of the
\emph{Hilbert series}~\cite{Lehman:2015via,Lehman:2015coa,Henning:2015daa,Henning:2017fpj}, which is a mathematical tool counting the number
of invariants in a graded algebra. For EFT purposes the Hilbert series 
\begin{equation}
H(\mathcal{D},\{\phi_i\}) = \sum_{k=0}^\infty \sum_{n_1=0}^\infty ... \sum_{n_N=0}^\infty c_{k,n_1,...,n_N} \mathcal{D}^k \phi_1^{n_1} ... \phi_N^{n_N}    
\end{equation}
counts the number $c_{k,r_1,...,r_N}$ of independent (higher-dimensional) operators involving $r_i$ fields $\phi_i$ and $k$ (covariant) derivatives, where the different variables act as labels and are simply used to count the number of occurrences
of the field in the operator (i.e. they are not the fields themselves).

In recent years, Hilbert series applications to EFT problems have proliferated,
and this paper aims to add a contribution to this body of literature. In this
section, we will recap some of the features and procedures involved, but we will
skip many of the more technical and mathematical aspects and refer the
interested reader to the literature. Instead, we will highlight the general
ideas, and lay the groundwork to implement discrete groups in the next section.

At its core the Hilbert series consists of two steps: It first provides a
generating function for all possible combinations of all available fields at a
given EFT order, and then projects out all those combinations which transform
under a given representation (typically the trivial one) of the symmetry group of the
theory. The main mathematical expressions needed are the \emph{characters} (or more specific, the eigenvalues) of
the group representations involved --- the characters are the traces of the group
elements, and are for Lie groups of rank $r$ given by functions of $r$ complex
variables. Characters of group representations exhibit two features crucial for
the Hilbert series: First, the character of a product of two representations
equals the sum of characters of the irreps contained in the product
representation. As an example, the Clebsch-Gordan decomposition each undergrad student encounters, which describes representations of SU(2)
\begin{equation}
\mathbf{2}\otimes\mathbf{2}=\mathbf{3}\oplus\mathbf{1}
\end{equation}
translates to a relation between characters $\chi_n(z)$ of the $n$-dimensional
representations\footnote{For a bit of intuition, look at the exponents on the variable $z$ in the
three characters involved. Up to a factor of two, these are $\pm\frac{1}{2}$
for the 2-dim irrep character, $\pm 1$ and 0 for the 3-dim one, and 0 for
the singlet, corresponding to the spin states or magnetic quantum numbers associated with the irreps.}
\begin{equation}
\label{eq:qm}
\chi_2(z)\cdot \chi_2(z)=\Bigl(\frac{1}{z}+z\Bigr)\cdot
\Bigl(\frac{1}{z}+z\Bigr) = \Bigl(\frac{1}{z^2}+1+z^2\Bigr)+ 1 =
\chi_3(z)+\chi_1(z).
\end{equation}
This allows us (modulo spin statistics covered below) to derive all possible
irreps that can be built from a set of field operators by simply multiplying
the characters of the fields.

We are then of course left with the task of identifying the correct irreps and
discarding those with the wrong transformation behaviour, for which we can employ the
second feature of characters: Two characters of irreps $R$ and $R'$ obey the
orthogonality relation
\begin{equation}
\int \df\mu_H \, \chi_R\,\chi_{R'}^{*}=\delta_{RR'},
\end{equation}
where $\mu_H$ is the Haar measure associated with the group in question. Concretely this is a contour integration over one or more complex variables, and for our purposes simply amounts to taking residues of polynomials. This
relation allows us to project out the number of singlets (or any other irrep)
contained in a product representation by simply replacing one of the two entries with the trivial character ``1'' (or whichever character we should use for any other target irrep we seek).

This rosy picture is complicated slightly by a few additional constraints
imposed by our target application: quantum fields obey (anti)commutation
rules, we discard operators related through the equations of motion, and we are
not interested in total derivatives. The latter two are not of any relevance for
the discrete symmetry cases we investigate in this paper as they relate
exclusively to the behaviour under the Lorentz group, and we refer the reader to
the relevant literature~\cite{Henning:2015daa,Henning:2017fpj}\footnote{As a quick recap, total derivatives are covered by using
representations of the conformal group of the Poincaré group, and the EOM relations are included by subtracting the
characters of the EOMs from the field characters.}.

This leaves the commutation behaviour: bosons commute, and fermions anticommute.
The question for us is therefore not whether e.g. the product representation of
two irreps given by some field contains a singlet, but
whether the \emph{symmetric} or \emph{antisymmetric} product representation
contains a singlet, where the symmetric case must be used for bosonic fields,
and the antisymmetric case for fermions. This generalises of course to higher
powers.
As a result we are not looking for a generating function for the characters of
arbitrary product representations, but for a generating function for the
characters of symmetric or alternating tensor powers, instead. This task can be
solved by taking a step back and remembering what representations and characters
actually are: Representations are matrices acting on vector spaces, and
characters are traces of group elements, i.e. they are sums of eigenvalues of
these matrices. As an example, take a 3-dimensional representation with
eigenbasis $e_1$, $e_2$, and $e_3$, and associated eigenvalues
$\lambda_i$\footnote{For e.g. the SU(2) adjoint we could use
$\lambda_1=z^{2}$, $\lambda_2=1$, $\lambda_3=z^{-2}$ with some $z$,
depending on the group element, as seen above. The SU(3) fundamental corresponds to $\lambda_1=z_1$, $\lambda_2=\frac{z_1}{z_2}$,
$\lambda_3=\frac{1}{z_2}$, which depends on two complex variables, as SU(3) is a rank-2 group.}.
The symmetric square of two such representations would have six eigenbasis vectors $e_1\otimes
e_1$, $e_2\otimes e_2$, $e_3\otimes e_3$, $e_1\otimes e_3+e_3\otimes e_1$,
$e_1\otimes e_2+e_2\otimes e_1$, and $e_2\otimes e_3+e_3\otimes e_2$. The
associated eigenvalues follow from the appearance of the $e_i$, and the
character for such a symmetric square representation $\chi_{Sym^{\otimes 2}}$ would thus
be
\begin{equation}
\chi_{Sym^{\otimes 2}}=\lambda_1^2+\lambda_2^2+\lambda_3^2+\lambda_1\lambda_2+\lambda_1\lambda_3+\lambda_2\lambda_3.
\end{equation}
Likewise, for the alternating square, the product eigenbasis is spanned by
$e_1\wedge e_2$, $e_1\wedge e_3$, and $e_2\wedge e_3$, and thus the alternating
square character $\chi_{\bigwedge^{\otimes 2}}$ is 
\begin{equation}
\chi_{\bigwedge^{\otimes 2}}=\lambda_1\lambda_2+\lambda_1\lambda_3+\lambda_2\lambda_3.
\end{equation}
We can identify these expressions as the complete homogeneous symmetric
polynomials (in the bosonic case) and elementary symmetric polynomials (in the
fermionic case) of degree 2 in 3 variables. This pattern can easily be extended,
and we can pattern-match the symmetric or alternating $m$th power of an
$n$-dimensional representation with the complete homogeneous or
elementary symmetric polynomials of degree $m$ in $n$ variables, where the
variables to be used are the eigenvalues of the group elements.
 
This excursion has an important point: There are generating functions for the
symmetric polynomials, the so-called \emph{plethystic exponentials}. Given a set of variables $x_1,\ldots,x_n$, the complete homogeneous symmetric polynomials of degree $k$ $h_k(\{x\})$ (in the $n$ variables $x_i$) obey 
\begin{equation}
    \sum_{k=0}^\infty h_k(\{x\}) t^k = \textrm{PE}\Bigl(\sum_{i=1}^n t\,x_i\Bigr) = \exp\Biggl(\sum_{i=1}^\infty\frac{t^i (x_1^i+\ldots + x_n^i)}{i}\Biggl),
\end{equation}
where we introduced a bookkeeping parameter $t$ to distinguish the various degrees. For the elementary symmetric polynomials $e_k(\{x\})$ a similar generating function exists:
\begin{equation}
    \sum_{k=0}^\infty e_k(\{x\}) t^k = \textrm{PEF}\Bigl(\sum_{i=1}^n t\,x_i\Bigr) = \exp\Biggl(-\sum_{i=1}^\infty \frac{(-t)^i (x_1^i+\ldots + x_n^i)}{i}\Biggl).
\end{equation}
We require bookkeeping parameters for the mass dimension $q$, which doubles as our EFT expansion parameter, as well as the labels of the fields $\varphi$, and the sums of (powers of) variables are given for our application by the characters $\chi_R(g^n)$, so that we use
\begin{align}
\begin{aligned}\label{eq:plethystic}
\textrm{PE}(q,\varphi,g)&=\exp\left(\sum_{k=1}^\infty\frac{(q^\Delta
\varphi)^k}{k}\chi_R(g^k)\right),\\
\textrm{PEF}(q,\varphi,g)&=\exp\left(-\sum_{k=1}^\infty\frac{(-q^\Delta
\varphi)^k}{k}\chi_R(g^k)\right),
\end{aligned} 
\end{align}
with $\textrm{PE}(q,\varphi,g)$ for bosons, and $\textrm{PEF}(q,\varphi,g)$ for fermions. $\Delta$ here is the mass dimension for the field in question (i.e. in 4 spacetime dimensions these are 1 for scalars, $\frac{3}{2}$ for spin-$\frac{1}{2}$ fermions, etc.), and $R$ labels the representation under which the field is assumed to transform.
Note here that the character for the symmetric or alternating powers involves the character of the naive power $\chi_R(g^k)$, which realistically means the sum of the $k$th powers of the eigenvalues of the representation matrix $R(g)$ for the group elements $g$.

Armed with this toolkit, the number of invariant operators for SMEFT-like theories can be derived
easily, e.g. by using a beautiful tool written in \texttt{FORM} by some of the authors of this paper. We will now look at the groups relevant for discrete flavour symmetries, and what difficulties they add.

\section{Discrete groups and the Hilbert series}
\label{sec:discrete}
The finite groups used in flavour physics are in many respects simpler than the
Lie groups treated before, and this is reflected in the Hilbert series. For
example, while the Lie groups require us to perform a complex integral involving
the Haar measure of the group (though in practice we must simply pick residues),
finite groups involve a sum over all group elements, which simplifies
--- as we are interested in characters, i.e. traces, which are class functions
--- into sums over conjugacy classes. In other words, to project out singlets
from some complicated product character $\Xi(g)$, dependent on the group
elements, we simply take
\begin{equation}
\#\text{singlets} =  \frac{1}{|G|}
\sum_{g\in G} \, 1 \cdot \Xi(g) = \frac{1}{|G|}
\sum_{k=1}^{N_c} n_k \bigl( 1 \cdot \Xi(g_k)\bigr).
\end{equation}
Here $|G|$ is the order of the group $G$, the second sum runs over all $N_c$
conjugacy classes, and $n_k$ is the number of elements in conjugacy class $k$. It follows that $\Xi$ involves the character(s) of a chosen class representative $g_k$ for each term in
the sum. However, the devil is in the details, and in particular the latter
statement is a problem. For the typical EFT applications we are still dealing
with plethystic exponentials here, which require the character of powers of group elements. We saw above that this
effectively means we require not the characters per se, but the eigenvalues of
the group elements, as the characters of powers of group elements are the sum of
powers of the eigenvalues. 

This poses somewhat of a problem, as typically one can find the characters in the literature, but not the individual eigenvalues that are summed up in the characters. For finite groups this is in particular a problem, as the eigenvalues are simply numbers, and so cancellations in the  character sums frequently occur. As an example a character for the $A_4$ triplet encountered below is the sum of the three third roots of unity, which vanishes. For Lie groups the problem is less acute, as there the characters are typically functions of variables, for which cancellations are not explicitly occurring, and so the eigenvalues can be read off: The $SU(2)$ adjoint representation has character $\chi_3(z)=z+1+\frac{1}{z}$, with each term one eigenvalue\footnote{The point is that for certain values of $z$ this may vanish or simplify, but not for generic $z$.}. 

We can in principle reconstruct the eigenvalues from the characters by using that they have to be complex numbers on the unit circle, and that the dimension of the representation tells us how many of them there are. For Lie groups in the context of \texttt{DECO} this is enough, as all representations we might encounter can be derived from applying $SU(2)$ lowering operators to a given highest weight state.
 This means the only non-variable-dependent constants we may encounter are integers, such as the ``1'' in the $SU(2)$ adjoint above, which arises from the mixed term $z \cdot \frac{1}{z}$ in the product of two doublets (see \eqref{eq:qm}) --- we get integers only if a variable and its \emph{own} conjugate collide. It is also evident that negative integers (or, in fact, sign flips in front of variables) cannot be constructed in this way, and so no terms can ever cancel. Any integer found in a character must therefore be the sum of multiple ``1''s --- see for example the $SU(3)$ adjoint, $\chi_8(z_1,z_2)=z_1 z_2+\frac{z_2^2}{z_1}+\frac{z_1^2}{z_2}+2 +\frac{z_2}{z_1^2}+\frac{z_1}{z_2^2}+\frac{1}{z_1z_2}$, which contains a ``2'' due to this pattern --- and multiples of monomials of variables are likewise positive sums of individual monomials of absolute value 1. 
 
 This latter property is ultimately what allows us to use eq.~\eqref{eq:plethystic}: As any non-variable eigenvalues are simply 1, we can get the character of some power of a group element by simply increasing the power of the variables appearing in the character. This is visibly true in e.g. the $SU(2)$ adjoint above, and can also be seen in e.g. the $SU(3)$ adjoint, which depends on two variables in various products and ratios, and contains a ``2'' which remains a ``2'' even when we consider the characters of powers of group elements. 
 
 However for finite groups even this entire mode of reconstruction can fail\footnote{Technically the reconstruction can also fail for Lie groups in wider contexts: In e.g. the case of the Hilbert Series with parity and/or charge conjugation we have to derive ``twining characters''\cite{Fuchs:1996vp,Fuchs:1996ju,Graf:2020yxt} arising from the action of $\mathcal{P}$ or $\mathcal{C}$. These can inject signs in unsavoury places, e.g. an element of the \emph{improper} orthochronous part of the Lorentz group for the $(\frac{1}{2},\frac{1}{2})$ representation can be diagonalised to yield eigenvalues $z,1,-1,z^{-1}$, with $z$ as always on the unit circle, and thus cancellations are back on the menu. However the treatment of parity and charge conjugation are not (or at least not yet) part of \texttt{DECO}.}: The tetraplet of $A_5$ for example contains a character of 0, which could arise from eigenvalues $1,1,-1,-1$, or $1,\text{i},-1,-\text{i}$. For finite groups it is therefore easiest to simply diagonalise the representation matrices explicitly.

\medskip

We will now apply this procedure to the group $A_4$, as a demonstration. $A_4$
has three 1-dimensional irreps, and one 3-dimensional one. The characters for
the 1-dimensional irreps are trivial, and not ambiguous. For the 3-dimensional
irrep, we must find the eigenvalues directly. The details of the group structure
of $A_4$ are listed in appendix~\ref{appendix}, we follow the notation of~\cite{Ishimori:2010au}, and we need to pick one
representative for the 3-dimensional irrep from each conjugacy class. Looking at
the distribution of elements in the classes, the matrices $a_1,a_2,b_1,c_1$ are sensible choices, with the first two already diagonal. As we only require the eigenvalues, the class representatives do not have to be diagonalised simultaneously, but we can diagonalise each matrix in turn (using different transformations), yielding
\begin{align}
\Lambda_{a_1}=\begin{pmatrix} 1\, & 0\, & 0 \\ 0\, & 1\, & 0  \\ 0\, & 0\,
& 1
\end{pmatrix}\quad\Lambda_{a_2}=\begin{pmatrix} 1\, & 0\, & 0 \\ 0\, & -1\, & 0  \\
0\, & 0\, & -1
\end{pmatrix} \quad
\Lambda_{b_1}=\begin{pmatrix} 1\, & 0\, & 0 \\ 0\, & \omega^2\, & 0  \\ 0\, & 0\,
& \omega
\end{pmatrix}\quad\Lambda_{c_1}=\begin{pmatrix} 1\, & 0\, & 0 \\ 0\, & \omega^2\,
& 0 \\ 0\, & 0\, & \omega
\end{pmatrix},
\end{align}
with $\omega:=e^{\img\frac{2\pi}{3}}$, from which the eigenvalues for the plethystic exponentials can be read off. 
A similar procedure can be performed for the elements of $S_4$, as laid out in the appendix.

Finally, a quick comment on the last family of discrete symmetries we wish to
consider, $\mathds{Z}_n$. This family of symmetries is nearly trivial, it
simply consists of the requirement that under the symmetry each field transforms
as $\varphi\rightarrow e^{\img\frac{2\pi k}{n}}\varphi$, where $k$ denotes the
charge of field $\varphi$.
Invariants do not pick up a phase, and in the Hilbert series we can implement this
by defining an auxiliary variable $\tau:=e^{\img\frac{2\pi}{n}}$. We equip each
occurrence of a field $\varphi$ with a $\tau^k$, imposing $\tau^n\equiv 1$ on
the Hilbert series, and then set $\tau=0$ to eliminate all operators
carrying residual charge under this symmetry.

\section[DECO]{\deco}
\label{sec:deco}
We are now in a position to lay out the design of our program. As is clear from the previous sections, we need the capability to deal with polynomials, arising from the expanded plethystic exponentials. Furthermore, the expressions we deal with are polynomials in variables corresponding to different groups, and the results of projecting out singlets under a given group will be derived by either summing over explicit values (for discrete groups) or picking residues (i.e. performing contour integrals for Lie groups). This will yield polynomials in the variables associated with the remaining groups.

Together, this draws the blueprint for a program: We write the program in \texttt{FORM}~\cite{Vermaseren:2000nd,Ruijl:2017dtg}, as it is the prime candidate for computations involving polynomials and their manipulation, and we exploit the polynomial structure to design a modular program, in which the output after one group's treatment is of the same form as the input. This means we can add  symmetry groups as we like, and treat them all in turn until there are none left to cover.

This leaves the question of which groups and representations we should add to the implementation. As we anticipate the main application to be BSM physics, we obviously allow for SU(3), SU(2), and U(1) groups, with trivial, defining, and adjoint representations. This allows us to employ the SM fields and representations, and enables any extensions based on gauged and ungauged versions of these groups, such as $Z'$ models or unitary flavour symmetries. The modular nature of \deco means that a user can add any number of these groups to the program, to accommodate even the most ambitious model builder's needs. In terms of discrete groups we add (for now) the groups $A_4$ and $S_4$ with their respective representations, as well as the $\mathds{Z}_n$ family. Again, multiple instances of each group can be defined. Finally, to accommodate low-energy supersymmetric models, we allow for U(1) symmetries with residual charge to account for the Grassmann integration measure's $R$-charge of -2 in $\mathcal{N}=1$ SUSY, which allows the EFT basis to carry the required compensating charge\footnote{The proximate reason we include this is due to the model presented in~\cite{Altarelli:2005yx}}.

For the fields that carry the charges under these groups we implement scalars, chiral fermions, Dirac fermions, field strength tensors, and Weyl tensors\footnote{These are the $(0,0)$, $(\frac{1}{2},0)$, $(0,\frac{1}{2})$, $(\frac{1}{2},0)\oplus(0,\frac{1}{2})$, $(1,0)\oplus(0,1)$, and $(2,0)\oplus(0,2)$ Lorentz representations.}. 

We wish to stress that this set of groups and representations does not exhaust the potential of \deco: Adding additional groups and/or representations requires only knowing the set of eigenvalues in each conjugacy class for each added representation of a group, and a small addition to the code implementing the replacement rules originating from the characters and Haar measure.

With these parameters, a typical \deco run proceeds according to a simple set of loops after the initial setup (more details can be found in the PDF manual accompanying the code, uploaded to \url{https://www.github.com/cbmarini/deco}):
\begin{enumerate}
\item The user specifies: 
\begin{itemize}
\item the mass dimension aimed for,
\item whether EOM and IBP redundancies should be considered,
\item the number of independent fermion generations,
\item and the names of the fields involved.
\end{itemize}
\item In the same file, all appearing fields are specified by type, and equipped with the various charges under all the groups. 
\item The program is then executed, and proceeds without further user input.
\end{enumerate}
Behind the scenes, the program runs through the following steps:
\begin{enumerate}
\item  The plethystic exponentials are expanded to the order needed for the specified EFT order. 
\item For each group type, the charges for one of the symmetries of this type are replaced with characters and residues are picked (or sums performed), and the program repeats this until all symmetries of the current group type have been processed.
\item This latter sequence is repeated for the other types of groups, with the order covered being Lorentz, then U(1), then SU(2), then SU(3), then $\mathds{Z}_n$, then $A_4$, then $S_4$\footnote{This corresponds to a comparatively quick execution of the program for a typical EFT, with the most restrictive groups (i.e. those with the capacity to reduce the number of terms the most, namely the Lorentz group, as well as all appearing U(1) groups) treated first.}.   
\item The result, a polynomial enumerating the effective Lagrangian (with field labels as the variables) is printed to the screen.
\end{enumerate}
A subtlety affects the treatment of $\mathds{Z}_n$ and U(1) with residue, as these families require two pieces of user-specified information: The member of the family (i.e. cycle length or residual charge), and the charge carried by a field itself. This can lead to ambiguity if the charges under these groups are not specified fully for all fields, as explained in the manual in more detail.

We devote one last comment to the name: while the moniker \deco might suggest otherwise, this program is no extension of the previously released \texttt{ECO}\footnote{\deco and \texttt{ECO} are in fact incompatible at the surface level already (and \texttt{ECO} files cannot be used with \deco), as their methods of defining the user input (dedicated procedures for \texttt{ECO}, a single sum of field entries for \deco) differ substantially.}~\cite{Marinissen:2020jmb}. In the latter we sacrificed flexibility for speed, and limited ourselves to only one SU(3) and SU(2) group each, which allowed us to structure the program to keep the number of expressions small even for EFT orders as high as 20 or more for SMEFT-like theories. The design application was to investigate the growth of operator numbers at high EFT orders. Here, we aim for flexibility, which means \deco will be significantly slower at higher EFT orders\footnote{e.g. at EFT order 6 in the SMEFT, \texttt{ECO} is a factor of 2 faster (0.11s for \deco vs 0.04s for \texttt{ECO} running on a single core of a run-of-the-mill laptop), while at EFT order 12, \texttt{ECO}'s advantage increases to a factor of 10 (50s for \deco vs. 5s for \texttt{ECO}). At mass dimension 14, \deco even fails to terminate within half an hour, while \texttt{ECO} wraps up within half a minute.}, but can accommodate combinations of groups \texttt{ECO} could not. The target problem for \deco is a quick estimate of what impact adding fields or symmetries to a Lagrangian has on the number of EFT operators at low subleading orders, and the capability of getting a swift sanity check on the number of operators used for phenomenological applications --- a tool to eliminate the possibility of over- \emph{and} undercomplete operator bases, even for theories with intricate symmetry structure. 

\section{Case studies}
\label{sec:models}
We now demonstrate how our newly programmed tool can be used for some explicit case
studies.

\smallskip

First, as a simple consistency check of our code we apply \deco to the SMEFT, and reproduce the cardinalities of the Warsaw basis~\cite{Grzadkowski:2010es} and dimension-8 basis~\cite{Murphy:2020rsh} as expected. To check the capability of using multiple U(1) symmetries we also verify the set of baryon number violating operators. Then, to test \deco against models with additional fields and multiple non-abelian symmetries of the same group type, we set out to check the number of subleading operators in the 2HDM- and MLRSM-EFT bases in~\cite{Anisha:2019nzx}, which we confirm up to an obvious typo in table 9 of that paper.

\smallskip

Finally, to test the implementation of the discrete group capability we set out to confirm the results of~\cite{Altarelli:2005yx},
or find any inconsistencies therein. This model (some further details of which can also be found in~\cite{Altarelli:2005yp}) postulates the usual Standard Model fields, as well as a modified scalar sector with several new
vev-acquiring scalar fields $\varphi$ and $\xi$ with various diacritics, sub-, and superscripts, plus a Higgs sector with two Higgs fields $h_u$ and $h_d$. 
In addition to the usual Standard Model symmetries, three additional symmetries
are introduced: 
\begin{itemize}
  \item An $A_4$ symmetry under which the lefthanded leptons transform as a
  triplet, and the charge-conjugate righthanded leptons transform under one of
  the three singlets, each. Some newly added fields of the scalar sector also transform non-trivially under this symmetry.
  \item A $\mathds{Z}_3$ symmetry, under which the leptons and some of the scalars are charged.
  \item A U(1)$_R$ symmetry, arising from supersymmetry, under which the leptons and some new ``driving'' scalars are charged. This symmetry is interesting mainly because all operators must carry
  residual charge 2, rather than being singlets, to make up for a charge of -2 carried by the supercoordinate measure. 
\end{itemize}

The complete set of fields and their charges is presented in table~\ref{AFcharges} below.

\begin{table}[htbp]
\centering
\renewcommand{\arraystretch}{1.3}
\begin{tabular}{c|cccc|ccccc|ccc}
%\hline
 & $l$ & $e^c$ & $\mu^c$ & $\tau^c$ & $h_{u,d}$ & $\phi_T$ & $\phi_S$ & $\xi$ & $\tilde{\xi}$ & $\phi_0^T$ & $\phi_0^S$ & $\xi_0$ \\ \hline\hline
$A_4$     & \textbf{3}             & \textbf{1}             & \textbf{1''}                    & \textbf{1'}                     & \textbf{1} & \textbf{3} & \textbf{3}        & \textbf{1}        & \textbf{1}        & \textbf{3} & \textbf{3}        & \textbf{1}        \\
$\mathds{Z}_3$     & $\omega$      & $\omega^2$    & $\omega^2$             & $\omega^2$             & 1 & 1 & $\omega$ & $\omega$ & $\omega$ & 1 & $\omega$ & $\omega$ \\
U(1)$_R$  & $\frac{1}{2}$             & $\frac{1}{2}$             & $\frac{1}{2}$                      & $\frac{1}{2}$                      & 0 & 0 & 0        & 0        & 0        & 1 & 1        & 1        \\ \hline
Mass Dim. & $\frac{3}{2}$ & $\frac{3}{2}$ & \textbf{$\frac{3}{2}$} & \textbf{$\frac{3}{2}$} & 1 & 1 & 1        & 1        & 1        & 2 & 2        & 2        %\\ \hline
\end{tabular}%
\caption{Field content of the model described in~\cite{Altarelli:2005yx}, copied alongside the notation. Fields present in the SM carry their usual SM charges, with both Higgs fields $h_u$ and $h_d$ carrying the charges of the SM Higgs. Note that some scalars have non-standard mass dimensions.}\label{AFcharges}
\end{table}
Armed with this assignment of charges, we can set out to rederive the
statements of~\cite{Altarelli:2005yx}. First, we can look at the Yukawa sector at leading order, and find the
same number and field content of the operators listed. Second, we confirm the appearance of only one operator (per generation)
affecting the charged lepton mass matrix. While we cannot make direct statements about the
precise index structure of the operator, the fact that there is only one precludes any
possibility of ambiguity.

We now look at the neutrino mass matrix. The \deco output here indicates that
there are a number of different operators contributing, while the authors
of~\cite{Altarelli:2005yx} dispense with the operators themselves, and instead
discuss their impact post-SSB: all but three operators are claimed to reproduce
the mass pattern of the Yukawa terms. These three operators are (stripped of
constants and charge conjugations)
\begin{equation}\label{eq:AFops}
(\varphi_T\varphi_S)'(l h_u l h_u)''\quad\quad (\varphi_T\varphi_S)''(l h_u l
h_u)'\quad\quad \xi (\varphi_T (lh_u lh_u)_{3S}),
\end{equation}
where $()$, $()'$, $()''$, and $()_{3S}$ correspond to the three different
singlet, as well as the symmetric triplet combinations of two $A_4$
triplets\footnote{The triplets here are $\varphi_T$, $\varphi_S$, as well as
$l$ (or the SU(2) singlet $lh_u$, if you prefer).}.

The \texttt{DECO} output by contrast reads
\begin{equation}
\label{eq:HS}
5l^2 h_u^2 \varphi_S \varphi_T+ l^2 h_u^2 \xi \varphi_T.
\end{equation}
Remembering that the Hilbert series doesn't know about index structures, we
identify the operator $\xi (\varphi_T (lh_u lh_u)_{3S})$ with the second
expression of~\eqref{eq:HS}. Thus, we would expect that
three of the five operators in the first expression of~\eqref{eq:HS} should reproduce the Yukawa terms when we plug in vevs, and the remaining two operators correspond to the other two non-trivial operators identified in~\eqref{eq:AFops}, in order to reproduce the statements made in~\cite{Altarelli:2005yx}. 

At this point we have no choice but to derive the operators explicitly, and
check for linear independence. Fortunately we now know how many there should be! It is straightforward to establish that 
\begin{gather}
(\varphi_T\varphi_S)'(l h_u l h_u)''\quad\quad(\varphi_T\varphi_S)''(l h_u l
h_u)'\quad\quad (\varphi_T\varphi_S)(l h_u l h_u)\nonumber \\
((\varphi_T\varphi_S)_{3S}(l h_u l h_u)_{3S})\quad\quad((\varphi_T\varphi_S)_{3A}(lh_u l h_u)_{3S}),
\end{gather}
where $()_{3A}$ is an antisymmetric triplet combination,
are five operators which are linearly independent and singlets under
$A_4$. The first two operators can identically be found in~\eqref{eq:AFops}, for the other three we must plug in vevs and check
whether they reproduce the Yukawa mass patterns as expected. Performing the substitution, we
find that while the third and fourth operator indeed reproduce existing Yukawa patterns\footnote{The operator  $(\varphi_T\varphi_S)(l h_u l h_u)$ yields a contribution to the Yukawa pattern present via the leading order operator $\xi(l h_u l h_u)$. As $\langle(\varphi_S,\varphi_T)_{3S}\rangle$ yields a linear combination of $\langle\varphi_T\rangle$ and $\langle\varphi_S\rangle$, the operator $((\varphi_T\varphi_S)_{3S}(l h_u l h_u)_{3S})$ contributes to the patterns introduced by the operators $\xi (\varphi_T (lh_u lh_u)_{3S})$ and $(\varphi_S (lh_u lh_u)_{3S})$. The former is the unique subleading operator identified above, the latter a leading order Yukawa term.}, the operator $((\varphi_T\varphi_S)_{3A}(lh_u l h_u)_{3S})$, involving an antisymmetric triplet combination, 
contributes a new pattern, as listed in table~\ref{t:yuk}. We
conclude that the operator $((\varphi_T\varphi_S)_{3A}(lh_u l h_u)_{3S})$ is missing from the basis discussed in~\cite{Altarelli:2005yx}.

\begin{table}[htbp]
\centering
\renewcommand{\arraystretch}{1.3}
\begin{tabular}{l|l|l}
Operator & Order & Mass pattern \\
\hline\hline
 $(l h_u l h_u)$ & Leading & $\nu_e^2+2\nu_\mu\nu_\tau$\\
 $(\varphi_S (l h_u l h_u)_{3S})$ & Leading &  $\nu_e^2+\nu_\mu^2+\nu_\tau^2-\nu_e\nu_\mu-\nu_\mu\nu_\tau-\nu_\tau\nu_e$\\ \hline
 $\xi(\varphi_T(l h_u l h_u)_{3S})$ & Subleading$\quad\quad$ & $\nu_e^2-\nu_\mu\nu_\tau$\\
$(\varphi_T\varphi_S)'(l h_u l h_u)''\quad\quad$& Subleading & $\nu_\mu^2+2\nu_e\nu_\tau$\\
$(\varphi_T\varphi_S)''(l h_u l h_u)'$& Subleading & $\nu_\tau^2+2\nu_e\nu_\mu$\\
\hline
$((\varphi_T\varphi_S)_{3A}(lh_u l h_u)_{3S})$  & Subleading & $\nu_\mu^2+\nu_e\nu_\mu-\nu_\tau^2-\nu_e\nu_\tau$
\end{tabular}%
\caption{The mass patterns introduced by the various appearing operators. The first two are leading order Yukawa operators, the second group of three corresponds to the new subleading patterns found in~\cite{Altarelli:2005yx}, and the last operator is the one missing. We use simplified notation omitting charge conjugation matrices and transposes, to highlight the flavour pattern.}\label{t:yuk}
\end{table}

\medskip

So what does the operator we just found actually do, and can it topple any of the
conclusions of~\cite{Altarelli:2005yx}? 

Prima facie, the additional operator contributes independently to physical observables, such as cross sections, decay rates, or measurable quantities such as particle masses, and can be disentangled from the effects of other operators by suitably chosen measurements. The authors of~\cite{Altarelli:2005yx} however specifically investigate the effects of the subleading operators they identified onto the neutrino mass matrix post-SSB, to find out how these can distort the pattern of masses and mixings away from the leading order structure, out of concern whether subleading order effects might be non-negligible. We will thus follow, and not consider other effects (such as e.g. the additional operator's contributions to neutrino-Higgs scattering cross sections). 

To investigate the impact on the mass matrix, a set of parameters $\delta z_i$ is introduced in~\cite{Altarelli:2005yx}, which measure the deviations from the leading order, by defining
\begin{equation}
m_\nu^\text{(NLO)}=m_\nu^\text{(LO)}+\frac{v_u^2}{\Lambda}\begin{pmatrix} 
\frac{2}{3}\delta z_1 & -\frac{1}{3}\delta z_3 +\delta z_5 & -\frac{1}{3}\delta z_2 +\delta z_4 \\
-\frac{1}{3}\delta z_3 +\delta z_5 & \frac{2}{3}\delta z_2 +\delta z_4& -\frac{1}{3}\delta z_1\\
-\frac{1}{3}\delta z_2 +\delta z_4 & -\frac{1}{3}\delta z_1& \frac{2}{3}\delta z_3 +\delta z_5
\end{pmatrix},
\end{equation}
where $v_u=\langle h_u\rangle$, and $\Lambda$ is the scale of New Physics.
Additional operators, or changes to the vevs for $\varphi_T$ and $\varphi_S$ away from their leading order alignment, induce non-zero values for these parameters $\delta z_i$. Equipping our new operator with a Wilson coefficient $x_f$ in line with eq.~(32) of~\cite{Altarelli:2005yx}, we find that it induces additional contributions to the $\delta z_2$ and $\delta z_3$ parameters:
\begin{equation}
\delta z_2 = \delta z_2^\text{(AF)} + x_f \frac{v_T v_S}{\Lambda^2},\quad \quad \delta z_3 = \delta z_3^\text{(AF)} - x_f \frac{v_T v_S}{\Lambda^2},
\end{equation}
where the $\delta z_i^\text{(AF)}$ are the values already given in~\cite{Altarelli:2005yx} (the $v_i$ are the vevs of the $\varphi_i$).

The neutrino masses (arising from diagonalisation) are found in~\cite{Altarelli:2005yx} to be dependent on $\delta z_2$ and $\delta z_3$ exclusively through the sum $\delta z_2+\delta z_3$, where the contribution from the new operator drops out. However, in the lepton mixing matrix, things are different: While the mixing angle $\tan^2 \theta_{12}$ again only depends on $\delta z_2+\delta z_3$, the mixing parameters $\tan^2 \theta_{23}$ and $|U_{e3}|$ depend on $\delta z_2-\delta z_3$, where the new operator yields a contribution (these parameters also depend in rather intricate ways on the leading order operator coefficients and vevs). We wish to point out, however, that both of these mixing parameters receive contributions from the two subleading operators already found in~\cite{Altarelli:2005yx}, also through non-zero values for the $\delta z_i$. Any argument suitable to justify the smallness of subleading corrections that applies to these operators will thus, \emph{ceteris paribus}, apply to the contribution from $x_f$. 

For modelbuilding purposes, the additional operator may be of interest, however, as the contributions to $\delta z_{2,3}$ already listed in~\cite{Altarelli:2005yx} arise from subleading corrections to the vacuum alignment. As such they represent a trickle-down effect from changes to the bosonic sector of the theory, specifically the scalar self-interactions. The new operator modifies the fermion sector directly, and without changes to the scalar potential. 

Finally, we can take a look at the ``driving potential'' of~\cite{Altarelli:2005yx},
which we confirm up to small and obvious typoes.

\section{Conclusions}
\label{sec:conc}
We presented in this paper the first instantiation of the Hilbert series applied to discrete symmetries in code, intended to quickly and flexibly provide guidance on the structure and size of subleading operator bases for generic EFTs for BSM physics. The result, called \deco, is made available at \url{https://www.github.com/cbmarini/deco} for use by the community, and can enumerate subleading operator bases for effective theories with freely chosen field content and symmetries. The available fields, symmetries, and representations already encompass the usual SM symmetries and fields, as well as a handful of discrete symmetries, most notably $A_4$, $S_4$, and $\mathds{Z}_n$, as well as $U(1)_R$. Our code is highly modular and can easily be adapted to allow for additional symmetry groups and representations, should the demand arise.

We used \deco to check the subleading operator bases of various models in the wider literature, found a missing operator in a widely used and cited model, and investigated its effect on measurable observables. 

In an era of increased reliance on EFTs, and with the number of potential models with additional fields and symmetries essentially infinite, we are thus convinced (and attempted to demonstrate this in action) that \deco can be of tremendous value to any researcher looking for a tool to swiftly assist in the process of finding subleading operator bases and double checking their results.

\acknowledgments

We thank Ferrucio Feruglio, Wouter Waalewijn, and Jim Talbert for valuable comments on the manuscript, and Jaco ter Hoeve for assistance with Github. 
R.R. is supported by the Royal Society through grant URF\textbackslash R1\textbackslash 201500, and acknowledges prior support via the NWO projectruimte 680-91-122.

\appendix
\section[A4 and S4]{$A_4$ and $S_4$}
\label{appendix}
In this appendix we follow mostly the approach and notation of~\cite{Ishimori:2010au}, with some added bit of information to make the discussion more accessible.

\subsection[S4]{$S_4$}

$S_4$ is the group of all permutations of a set of four items, and thus contains $4!=24$ elements. Its conjugacy classes are determined by the cycle type, and so correspond to the set of unordered integer partitions of the number 4. There are thus five conjugacy classes corresponding to the partitions $4,3+1,2+2,2+1+1,1+1+1+1$, which also means due to orthogonality that it has five irreducible representations, which must be two singlets, a doublet, and two triplets. In terms of the permutations the conjugacy classes correspond to the identity (``1+1+1+1'', keeping all elements fixed), 6 4-cycles (``4'', three choices for the initial replacement, two for the second) , 8 3-cycles (``3+1'', with four possible fixed elements, cycling ``clockwise'' or ``anticlockwise''), 6 2-cycles (``2+1+1'', two out of  four elements swapped, the others left in place) and 3 double 2-cycles (``2+2'', the three ways of splitting a group of 4 elements in half to each be swapped). It follows that the classes 2+1+1 and 4 consist of odd permutations, the others of even ones.
\medskip

The two singlet representations are easy to deduce: The trivial representation and a sign representation which sends even permutations to 1, and odd ones to -1. For us this simply means that since we need one element out of every conjugacy class for the Hilbert series, we have three classes which get assigned a 1, and two which can be assigned a 1 or a -1, and we know their respective cardinality.

\smallskip

The doublet and triplet irreps require more work:
We can get the triplet via the construction of the \emph{standard representation} common to all groups $S_n$, which is ($n-1$)-dimensional, i.e. in this case a triplet:
We use that for a vector space spanned by four basis vectors $e_i$ with coordinates $x_i$ the sum of all coordinates $x_1+x_2+x_3+x_4$ is invariant under permutations, i.e. a trivial singlet. We can then build a matrix representation on the 3-dimensional vector space orthogonal to the vector $x_i e_i$\footnote{An alternative contruction would be to restrict ourselves to the subspace spanned by the constraint $\sum_i x_i=0$, which is equivalent.}. A list of the matrices for this representation can be found in~\cite{Ishimori:2010au}, and we only require one matrix from each conjugacy class for the Hilbert series (or more specifically we need their eigenvalues). A sensible choice would be the matrices $a_1,a_2,b_1,d_1,d_3$, whose diagonalised versions are

\begin{equation}
\begin{aligned}
& \Lambda_{a_1}=\begin{pmatrix} 1\, & 0\, & 0 \\ 0\, & 1\, & 0  \\ 0\, & 0\,
& 1
\end{pmatrix}\quad\Lambda_{a_2}=\begin{pmatrix} 1\, & 0\, & 0 \\ 0\, & -1\, & 0  \\
0\, & 0\, & -1
\end{pmatrix} \quad
\Lambda_{d_1}=\begin{pmatrix} -1\, & 0\, & 0 \\ 0\, & 1\, & 0  \\ 0\, & 0\,
& 1
\end{pmatrix} \\
& \qquad \qquad \Lambda_{b_1}=\begin{pmatrix} 1\, & 0\, & 0 \\ 0\, & \omega^2\,
& 0 \\ 0\, & 0\, & \omega
\end{pmatrix}\quad\Lambda_{d_3}=\begin{pmatrix} -1\, & 0\, & 0 \\ 0\, & \img\,
& 0 \\ 0\, & 0\, & -\img
\end{pmatrix},
\end{aligned}
\end{equation}
with $\omega=\exp{\img\frac{2\pi }{3}}$.

The second triplet can be found by applying the logic of the sign representation to the triplet case, by simply equipping every odd permutation (i.e. the matrices $d_1$ and $d_3$) with an overall minus.

\smallskip

Finally, the doublet. Here we can use that the Klein group $V$ (which is the union of the double 2-cycle conjugacy class and the identity) is a normal subgroup of $S_4$, the quotient $S_4/V\cong S_3$, and irreps of quotient groups with normal subgroups are in 1-to-1 relation with irreps of the original group. This means the standard irrep of $S_3$, which is a doublet, is an irrep
of $S_4$.
The construction can again be found in~\cite{Ishimori:2010au}, and as the quotient involves the double 2-cycle (which establishes an entire conjugacy class and relates through its action the 4-cycles and 2-cycles) the matrices for the double 2-cycles must match the identity, and the matrices for the 4-cycle must match those of the 2-cycle. We thus use
\begin{equation}
\begin{aligned}
& \Lambda_{a_1}=\begin{pmatrix} 1\, & 0\, \\ 0\, & 1
\end{pmatrix}\quad\Lambda_{a_2}=\begin{pmatrix} 1\, & 0\, \\ 0\, & 1
\end{pmatrix} \quad
\Lambda_{d_1}=\begin{pmatrix} -1\, & 0\, \\ 0\, & 1
\end{pmatrix} \\
& \qquad \Lambda_{b_1}=\begin{pmatrix} \omega\, & 0\, \\ 0\, & \omega^2
\end{pmatrix}\quad\Lambda_{d_3}=\begin{pmatrix} -1\, & 0\, \\ 0\, & 1
\end{pmatrix},
\end{aligned}
\end{equation}
which provides all the input needed for the Hilbert series.

\subsection[A4]{$A_4$}

$A_4$ is the group of even permutations on a set of four elements (and the symmetry group of a tetrahedron), thus a subgroup of $S_4$, which means a lot of properties carry over. In particular the conjugacy classes can be (initially) adapted, as we can take the classes of $S_4$ and discard the ones with odd permutations. We then simply need to find out if the remaining classes split, or carry over unchanged. They are respectively the class containing the identity (which obviously won't split into two classes), as well as the classes with 3-cycles and the double 2-cycle. The latter cannot split, as conjugation with the 3-cycles map the distinct double 2-cycles to each other, but the 3-cycles split, as the chirality of the 3-cycles is preserved under any conjugation. It thus splits into two classes with 4 elements each (1 out of 4 elements held fixed, and the rest cycled clockwise for one class, and anticlockwise for the other).

\bigskip

For the representations we find that $A_4$ must have four irreps, a triplet and three singlets. We can identify a trivial singlet as before, and we can generate the two non-trivial singlets by equipping the elements of the split conjugacy class with a phase of $\omega=\exp{\img\frac{2\pi}{3}}$ or $\omega^2$. The two choices for this assignment correspond to the two remaining singlets\footnote{Note here that the action of the identity and the double 2-cycles preserve the two 3-cycle classes. They should thus be represented by 1, which is trivial for the identity, but less so for the double 2-cycles.}. 

Finally again, the triplet. We can simply identify its matrices with their relevant counterparts from the $S_4$ triplet (and its $\mathbf{3'}$, as the odd permutations are the only difference between the two and are simply missing here).

Taking into account that the 3-cycle class splits, we use the diagonalised matrices
\begin{align}
\Lambda_{a_1}=\begin{pmatrix} 1\, & 0\, & 0 \\ 0\, & 1\, & 0  \\ 0\, & 0\,
& 1
\end{pmatrix}\quad\Lambda_{a_2}=\begin{pmatrix} 1\, & 0\, & 0 \\ 0\, & -1\, & 0  \\
0\, & 0\, & -1
\end{pmatrix} \quad
\Lambda_{b_1}=\begin{pmatrix} 1\, & 0\, & 0 \\ 0\, & \omega^2\, & 0  \\ 0\, & 0\,
& \omega
\end{pmatrix}\quad\Lambda_{c_1}=\begin{pmatrix} 1\, & 0\, & 0 \\ 0\, & \omega^2\,
& 0 \\ 0\, & 0\, & \omega
\end{pmatrix},
\end{align}
from which the eigenvalues for the plethystic exponentials can be read off.

\bibliography{bibliography.bib}

\end{document}